\documentclass[aps,prr,amsmath,amssymb,floatfix,twocolumn]{revtex4-1}

\usepackage{multirow}
\usepackage{bbold}
\usepackage{graphicx}
\usepackage{subfigure}
\usepackage{color}
\usepackage{hyperref}
\usepackage[justification=raggedright]{caption}
\usepackage{filecontents}

\begin{document}

\preprint{APS/123-QED}
\title{Evidence for topological origin of large spin-shift current in antiferromagnetic Ti$_{4}$C$_{3}$}
\author{Ali Sufyan$^{3}$}
\author{Hasan M. Abdullah$^{4}$}
\author{J. Andreas Larsson$^{3,5}$}
\author{Alexander C. Tyner$^{1,2}$}
\email{alexander.tyner@su.se}

\affiliation{$^{1}$ Nordita, KTH Royal Institute of Technology and Stockholm University 106 91 Stockholm, Sweden}
\affiliation{$^2$ Department of Physics, University of Connecticut, Storrs, Connecticut 06269, USA}

\affiliation{$^{3}$ Applied Physics, Division of Materials Science, Department of Engineering Sciences and Mathematics,
Lule\aa\,University of Technology, Lule\aa\, SE-97187, Sweden}

\affiliation{$^{4}$ Department of Physics, Faculty of Applied Science, Taiz University, Taiz, Yemen}

\affiliation{$^{5}$ Wallenberg Initiative Materials Science for Sustainability,
Lule\aa\,University of Technology, Lule\aa\, SE-97187, Sweden}

\date{\today}

\begin{abstract} 
The shift current is a non-linear photocurrent generally associated with the underlying quantum geometry. However, a topological origin for the shift photocurrent in non-centrosymmetric systems has recently been proposed. The corresponding topological classification goes beyond the ten-fold paradigm and is associated with the presence of a reverting Thouless pump (RTP). In this work we examine an antiferromagnetic monolayer within the family of MXenes, Ti$_{4}$C$_{3}$. This material is centrosymmetric, however, magnetic ordering violates inversion symmetry. We demonstrate evidence of an RTP in each spin-sector which has been perturbed, destroying quantization of the invariant. Nevertheless, a giant spin-resolved shift current persists. We further investigate the mid-gap edge states and classification of the system as a fragile topological insulator to which trivial bands have been coupled. 
\end{abstract}

\maketitle

\par 
\section{Introduction}
The pursuit of sustainable energy necessitates the development of cost-effective, high-performance solar cell technologies. While conventional photovoltaic devices, based on p-n junction architectures, have achieved considerable progress, their power conversion efficiency is fundamentally constrained by the Shockley-Queisser limit\cite{shockley2018detailed,markvart2022shockley}. This inherent limitation has motivated extensive research into alternative photocurrent generation mechanisms, among which the shift current effect has emerged as a particularly compelling candidate \cite{photo-1,photo-2,photo-3,photo-4,photo-5,photo-6}. Distinct from traditional photovoltaics, which rely on built-in electric fields at p-n junctions to facilitate charge carrier separation, the shift current is an intrinsic bulk phenomenon arising in non-centrosymmetric materials\cite{inversion-bulk-1,inversion-bulk-2,inversion-bulk-3,inversion-bulk-4,inversion-bulk-5,inversion-bulk-6,inversion-bulk-7,inversion-bulk-8,inversion-bulk-9}. This characteristic eliminates the requirement for a p-n junction, offering a potential pathway to circumvent the efficiency constraints imposed on conventional solar cells\cite{photo-4}. Furthermore, the photocarriers generated via the shift current mechanism have been shown to exhibit remarkably long transport distances, significantly enhancing their potential for efficient energy conversion\cite{dai2023recent, dai2021first, sauer2023shift}. This extended transport length minimizes recombination losses and allows for the collection of carriers generated far from the contacts, contributing to improved device performance.
\begin{figure}[h]
\includegraphics[width=8cm]{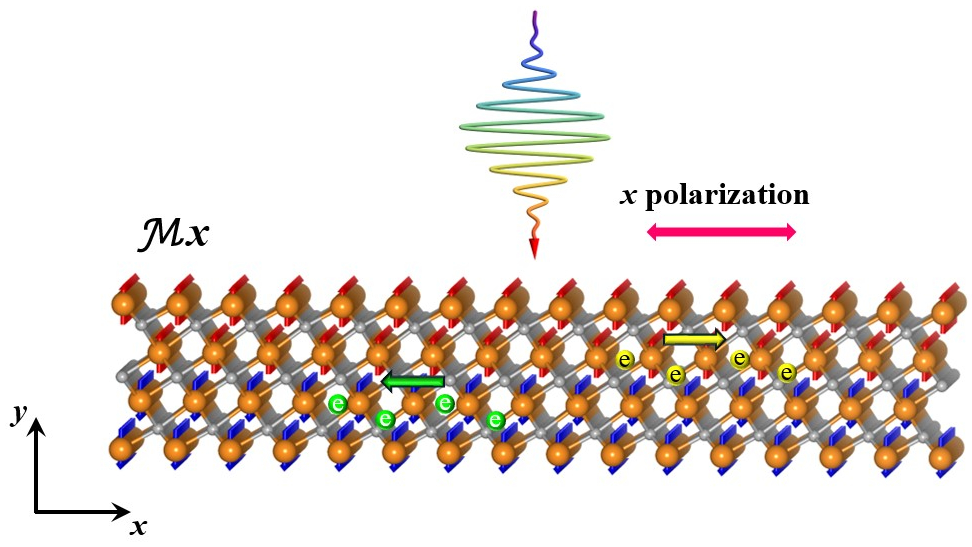}
\caption{Schematic detailing production of spin shift-current in Ti$_{4}$C$_{3}$ from incident, $x$-polarized light.}
\label{fig:fig0}
\end{figure}
\begin{figure*}[!t]
\includegraphics[width=5 in]{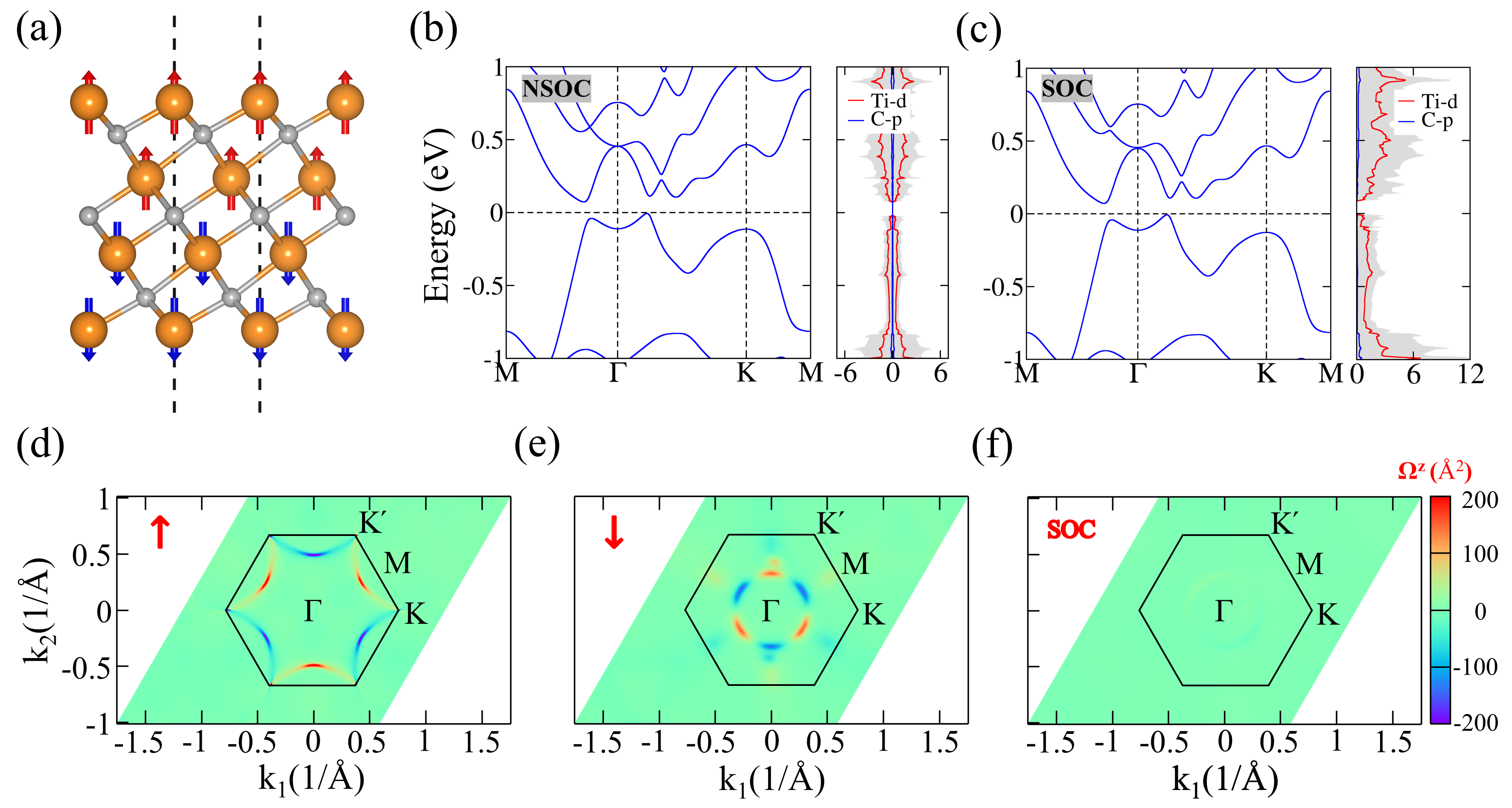}
\caption{(a) Crystal structure of monolayer Ti$_4$C$_3$ with AFM1 magnetic ordering. (b) Electronic band structures calculated without and (c) with spin–orbit coupling (SOC). (d) Berry curvature distributions in momentum space for spin-up, (e) spin-down, and (f) SOC-included cases, revealing spin-dependent and SOC-driven contributions}
\label{fig:fig1}
\end{figure*}
The shift current is a second$-$order nonlinear optical response that generates a direct current (DC) under continuous wave illumination in materials lacking inversion symmetry\cite{PhysRevB.61.5337, photo-4}. This phenomenon arises from the asymmetric distribution of electronic wavefunctions in real space, leading to a displacement of the charge centroid upon photoexcitation. The resulting real-space charge separation creates a persistent current even in the absence of an external bias. The shift current responses have been reported across a broad spectrum of material systems, including bulk perovskite ferroelectrics\cite{ferroelectric-1,ferroelectric-2,ferroelectric-3,ferroelectric-4,inversion-bulk-9}, two-dimensional (2D) materials\cite{2D-1,2D-2,2D-3,2D-4,2D-5,2D-6}, nanotubes\cite{nano-tube}, conjugated polymers\cite{polymer}, and topological materials\cite{topological-1,topological-2,topological-3,topological-4,topological-5}. The versatility of the shift current effect across diverse material classes underscores its fundamental nature and opens up a wide range of possibilities for material optimization. Recent theoretical and experimental advances have further elucidated the crucial role of excitonic effects in enhancing second-order optical responses, leading to a substantial increase in shift currents, particularly near resonance\cite{resonance-1,resonance-2}. This enhancement, attributed to inter-exciton coupling and resonant enhancement of the transition dipole moments, is significantly stronger than the corresponding increase in linear optical absorption observed near the band gap\cite{resonance-1,resonance-2,dai2021first,pedersen2015intraband}. This disparity suggests new and highly effective strategies for optimizing shift current materials by engineering excitonic properties. 

Importantly, while conventional shift current materials exhibit a geometric phase contribution that is not quantized, recent theoretical developments have identified a new class of topological insulators capable of supporting quantized shift currents\cite{alexandradinata2024quantization}. This quantization stems from the intricate interplay between intraband and interband Berry phases, resulting in a shift vector that is topologically protected and cannot be continuously deformed to zero. Unlike ordinary shift current materials, these systems possess a fundamentally robust nonlinear optical response, paving the way for novel and efficient optoelectronic applications. However, despite these theoretical predictions, a definitive experimental observation of a quantized shift current in a real material has remained elusive.

\begin{table}[h]
\centering
\caption{Total energy per unit cell calculated for different magnetic configurations.}
\begin{tabular}{|c|c|c|c|c|}
\hline
\begin{tabular}{@{}c@{}}NM \\ (eV)\end{tabular} &
\begin{tabular}{@{}c@{}}FM \\ (eV)\end{tabular} &
\begin{tabular}{@{}c@{}}AFM1 \\ (eV)\end{tabular} &
\begin{tabular}{@{}c@{}}AMF2 \\ (eV)\end{tabular} &
\begin{tabular}{@{}c@{}}AMF3 \\ (eV)\end{tabular} \\
\hline
-62.552 & -62.675 & -62.685 & -62.675& -62.684\\
\hline
\end{tabular}
\label{tab:tab1}
\end{table}

In this work, we investigate the possibility of a topological origin for the shift current in an antiferromagnetic monolayer MXene, Ti$_4$C$_3$ which has been recently synthesized\cite {ti4c3}. While structurally centrosymmetric, the AFM ordering in Ti$_4$C$_3$ breaks inversion symmetry, enabling a nontrivial shift current in it. The band topology is classified via identification of a reverting Thouless pump (RTP) for both spin-up and spin-down bands. To further examine its topological nature, we perform calculations of edge states, and Berry curvature. Our results establish, to the best of our knowledge, monolayer Ti$_4$C$_3$ as the first centrosymmetric material exhibiting a shift current linked to RTP with fragile topology, highlighting a new paradigm in nonlinear transport induced by magnetic ordering.

\section{Computational and structural details of Ti$_{4}$C$_{3}$}

First-principles calculations were performed within the framework of density functional theory (DFT) as implemented in the Vienna ab initio simulation package (VASP) \cite{paw1,paw2,vasp}. Projector augmented-wave (PAW) potentials were used to describe the electron-ion interactions, and the Perdew-Burke-Ernzerhof (PBE) functional was employed for the exchange-correlation potential\cite{perdew1996generalized}. A plane-wave energy cutoff of 550 eV was used, and the Brillouin zone was sampled using a 21 $\times$ 21 $\times$ 1 Monkhorst-Pack k-point grid\cite{monkhorst1976special}. A vacuum layer of 20 \text{\AA} was included along the out-of-plane direction to minimize interactions between periodic images. Structural optimizations were performed until the forces on all atoms were less than 0.001 eV/\text{\AA}. To address the strong on-site Coulomb interactions of Ti-3d electrons, the GGA+U method was applied with U$_{eff}$ = 3 eV\cite{finazzi2008excess,sakhraoui2022electronic}. The tight-binding Hamiltonian was constructed using maximally localized Wannier functions\cite{marzari1997maximally}, and the corresponding topological properties were analyzed with the WannierTools package\cite{wu2018wanniertools}.

\begin{figure*}
\includegraphics[width=16cm]{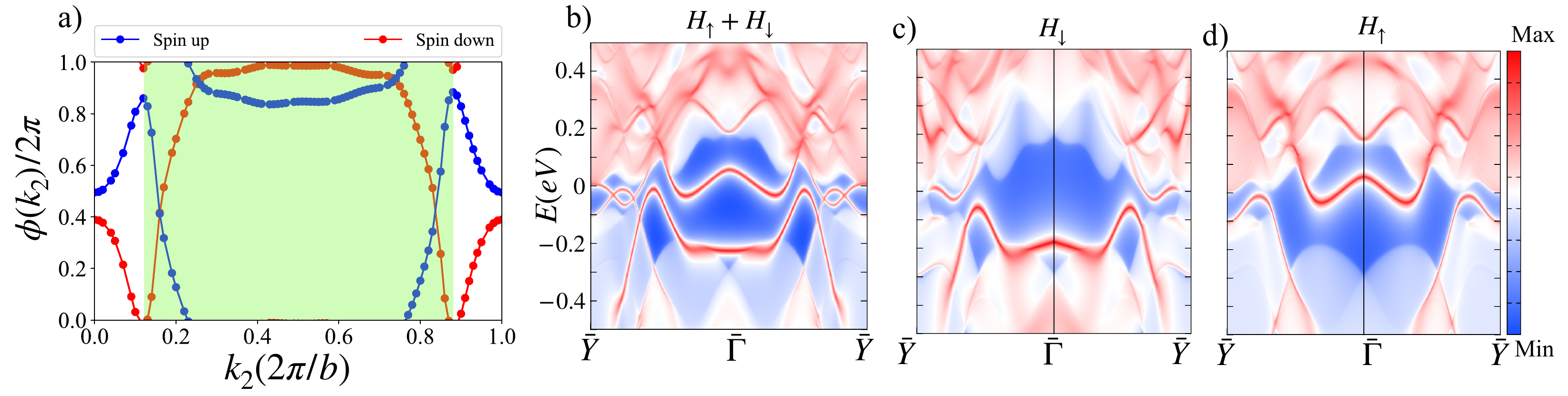}
\caption{(a) Berry phase accrued by the ground-state wavefunction upon parallel transport by a reciprocal lattice vector for the spin-up and spin-down sectors, as computed via eq. \eqref{eq:Berry}. (b)-(d) Surface spectral density on the (10) surface in the absence of spin-orbit coupling considering (b) the spin up and down sectors simultaneously, (c) only the spin-down sector and (d) only the spin-up sector. The surface spectra is computed using the WannierTools software package\cite{wu2018wanniertools}.}
\label{fig:fig2}
\end{figure*}

Monolayer Ti$_4$C$_3$ crystallizes in the trigonal space group P-3m1 with \textit{D}$_{3d}$ point group symmetry. The relaxed lattice parameters are \textit{a} = \textit{b} = 3.094 \text{\AA} and $\gamma$ = $120^\circ$. The structure exhibits a unique layered arrangement, featuring alternating Ti and C layers with a Ti-C-Ti stacking sequence. The structure exhibits inversion symmetry and three vertical mirror planes $\mathcal{M}$ perpendicular to the basal plane, along with a three-fold rotational axis (\textit{C}$_3$) along the \textit{c}-axis. A systematic investigation of magnetic configurations, including a ferromagnetic (FM) state and three different AFM orders (AFM1–AFM3), identifies AFM1 as the most stable configuration (see Fig. S1, Fig.\ref{fig:fig1} (a), and Table 1). The AFM1 state is characterized by antiparallel spin alignment between the upper two and lower two titanium layers. The observed magnetic moments exhibit a distinct spatial distribution within the Ti$_4$C$_3$ monolayer. The outermost Ti layers (top and bottom) lack direct bonding with C atoms, resulting in reduced hybridization and allowing for the formation of localized magnetic moments of approximately $\pm$0.78$\mu{_B}$. In contrast, the inner Ti layers, which are fully coordinated with carbon atoms, experience stronger hybridization effects, leading to a suppression of their magnetic moments to approximately $\pm$0.14$\mu{_B}$. Importantly, the AFM ordering in monolayer Ti$_4$C$_3$ breaks time-reversal symmetry ($\mathcal{T}$), as well as inversion symmetry ($\mathcal{P}$), which is structurally present in the paramagnetic state. However, the combined $\mathcal{P}$$\mathcal{T}$ symmetry is preserved in the AFM configuration. Figure \ref{fig:fig1}(b) illustrates the electronic band structure of Ti$_4$C$_3$, revealing its semiconducting nature with a small bandgap of 79 meV. Consistent with other MXene materials, the \textit{d}-states of Ti atoms dominate near the Fermi level, as highlighted by the partial density of states (PDOS). The inclusion of spin-orbit coupling (SOC), shown in Fig.\ref{fig:fig1}(c), reveals that every band remains at least doubly degenerate due to the protection of $\mathcal{P}$$\mathcal{T}$ symmetry. Notably, the band structure shows no significant changes with SOC, underscoring the weak SOC effect in Ti$_4$C$_3$.

The calculated Berry curvature distributions are presented in Fig.\ref{fig:fig1}(d-e). In the absence of SOC, the Berry curvature displays a pronounced asymmetry in momentum space, with regions of high positive and negative density distributed across the Brillouin zone. For spin-up and spin-down bands, the Berry curvature exhibits oppositely signed regions, resulting in significant cancellation when SOC is included. This cancellation effectively suppresses the linear quantum anomalous Hall effect in Ti$_4$C$_3$. 

\section{Non-linear response}
The presence of combined $\mathcal{PT}$ symmetry precludes the existence of a finite Chern number at the Fermi energy and corresponding finite anomalous Hall effect. Furthermore, the Chern number for the up and down spin sectors can be computed at the Fermi energy by tracking the Berry phase accumulated by the ground state wave function upon parallel transport by a single reciprocal lattice vector\cite{BerryPhase,Qi2008,YuZ2,ChernPolarization,VanderbiltWCC,z2pack,tyner2023berryeasy}. Explicitly, this Berry phase can be computed as, 
\begin{equation}\label{eq:Berry}
    \phi_{q\in\uparrow,\downarrow}(k_{2})=\sum_{n\in occ}\oint A_{1,n,q}(\mathbf{k})dk_{1},
\end{equation}
where $A_{1,n,q}(\mathbf{k})=-i\langle u_{n,q}|\partial_{k_{1}}| u_{n,q}\rangle$ where $u_{n,q}$ is the Bloch wavefunction of the $n$-th band labeled by spin $q$. The Chern number then follows as, $C=(\phi_{q}(2\pi/b)-\phi_{q}(0))/(2\pi)$. The results of this calculation are shown in Fig. \eqref{fig:fig2}(a) revealing that, despite the large Berry curvature density visible in Fig. \eqref{fig:fig1}(d)-(e) for the spin up and spin-down states independently, the net Berry flux is vanishing in each case. This result suggests that a quantized, topological contribution to the spin-Hall conductivity is absent. 
\par 
Despite the lack of an anomalous first-order response in the form of charge or spin-Hall conductivity, the large Berry curvature density in the spin-up and spin-down sectors combined with the presence of mid-gap edge states, seen in Fig. \eqref{fig:fig2}(b)-(d), motivates further investigation into topological classification beyond the ten-fold paradigm and non-linear response due to the underlying quantum geometry\cite{PhysRevB.23.5590,PhysRevB.52.14636,PhysRevB.61.5337,Sodemann2015,tokura2018nonreciprocal,cook2017design,du2021nonlinear,gao2023quantum,wang2023quantum,alexandradinata2024quantization}. In particular emphasis has been placed in recent years on the identification of materials supporting a large non-linear anomalous Hall effect (NLAHE)\cite{Sodemann2015} or shift current\cite{cook2017design} due to potential use in quantum technologies. Both the NLAHE and shift current are dependent on underlying symmetries of the system. Unfortunately, despite the large Berry curvature density, the net Berry dipole moment for Ti$_{4}$C$_{3}$ must be vanishing due to the presence of more than one mirror axis. 
We would additionally expect the shift current to vanish due to the presence of centrosymmetry however the anti-ferromagnetic ordering has the effect of breaking inversion symmetry, leading to the possibility that a shift current of equal and opposite magnitude can be identified for the up and down spin states, labeled a spin-shift current\cite{PhysRevB.95.035134}. 

\begin{figure}
\includegraphics[width=8cm]{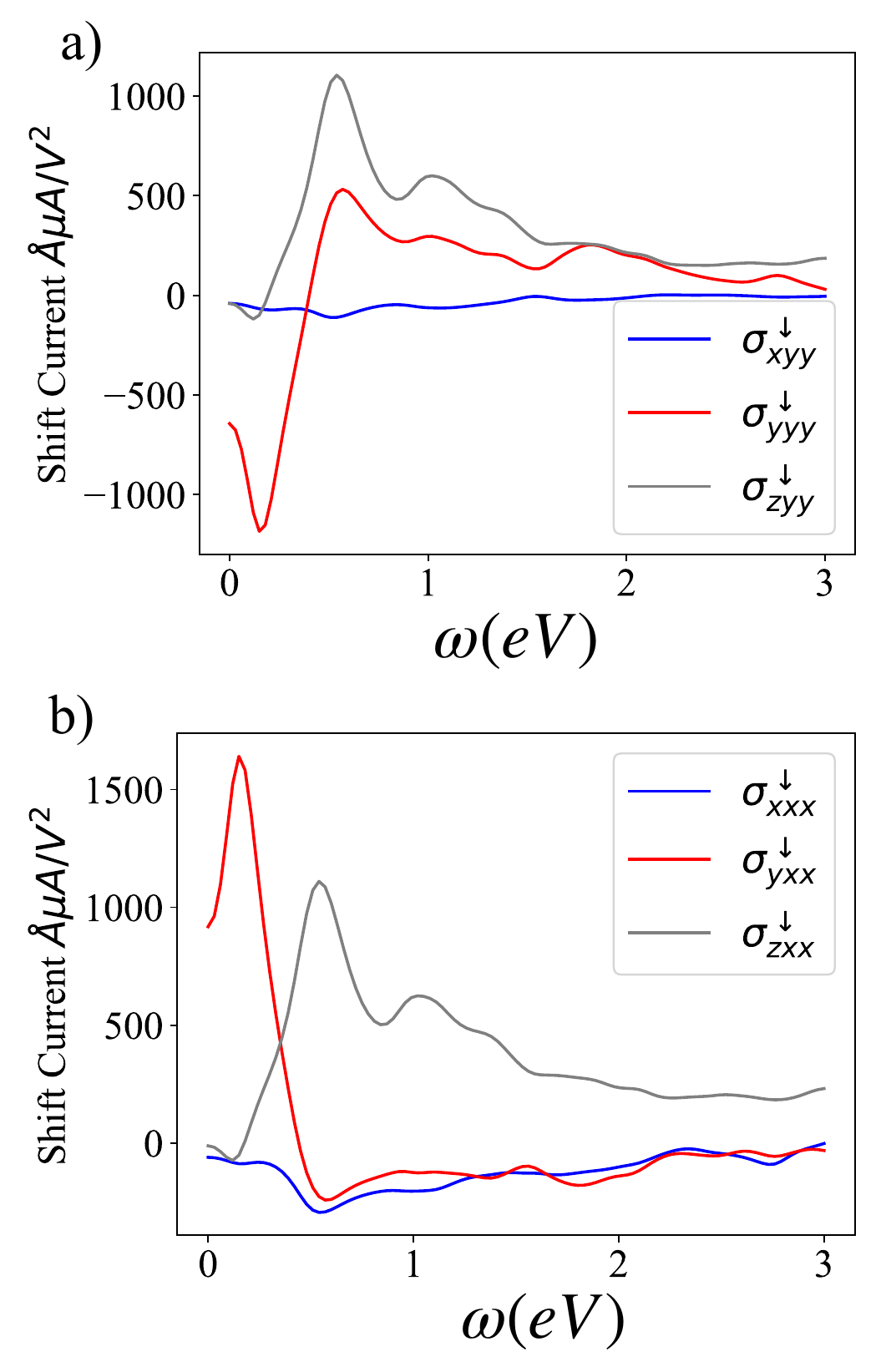}
\caption{ (a) Shift-current conductivity computed for light linearly polarized along the y-direction and (b) the x-direction.}
\label{fig:fig3}
\end{figure}
\par 
The shift current is one of two contributions to the total non-linear photo current, the injection current being the other,  and takes the form\cite{cook2017design}, 
\begin{equation}
    J_{2,SC}^{a}=\sigma_{abc}(0;\omega,-\omega)E_{b}(\omega)E_{c}(-\omega).
\end{equation}
The shift current conductivity tensor is defined as, 
\begin{multline}\label{eq:SC}
    \sigma_{abc}=-\frac{i\pi e^{3}}{2\hbar^{2}}\int \frac{d\mathbf{k}^{3}}{(2\pi)^{3}}\sum_{n,m}f_{nm}(r^{b}_{mn}r^{c}_{nm;a}+r^{c}_{mn}r^{b}_{nm;a})\\\times \delta(\epsilon_{mn}-\hbar \omega),
\end{multline}
where $m,n$ are band indices, $\epsilon_{mn}=\epsilon_{m}-\epsilon_{n}$ is the difference in band energy, and $f_{mn}=f_{m}-f_{n}$ is the difference in Fermi-Dirac occupation. We further define $r^{a}_{mn}=A^{a}_{mn}=i\langle m|\partial_{k_{a}} n\rangle$ is the interband Berry connection, and $r^{b}_{nm;a}=\partial_{k_{a}}r^{b}_{mn}-i(A^{a}_{nm}-A^{a}_{mm})r^{b}_{mn}$ is the generalized derivative of the interband Berry connection. In this work we are focused on linearly polarized light such that $b=c$ in eq. \eqref{eq:SC}. The shift current is computed using the WannierBerri software package\cite{tsirkin2021high}. In order to extract the shift current in two-dimensions the result is multiplied by the system size along the third-axis used in the DFT computations. 

The results are shown in Fig.  \eqref{fig:fig3}(a)-(b) considering polarized light along the $x$ and $y$ directions. Notably, multiple components of the shift current conductivity tensor display a large response when $\hbar \omega<2eV$. To place the magnitude of the shift current for each spin-sector in context, we remark that the maximum value of the shift current conductivity compares favorably with the best candidate compounds identified in Ref. \cite{sauer2023shift}; where a library of two-dimensional materials were screened for shift-current generation. Optimal compounds identified in Ref. \cite{sauer2023shift} displayed a maximal conductivity of 1-2$\AA mA/V^{2}$ for $\hbar\omega<3eV$; precisely in line with our computed response. 

While the large Berry curvature density seen in Fig. \eqref{fig:fig1}(d)-(e) can be used as an indication that the underlying quantum geometry is the source of the large effect, we can explore an alternative explanation following a recent work by Alexandradinata\cite{alexandradinata2024quantization}. In this work a topological origin of the shift current was introduced along with a novel topological invariant referred to as a reverting Thouless pump (RTP). The more familiar non-reverting Thouless pump is found by invoking eq. \eqref{eq:Berry} in the presence of a topological invariant such as a Chern number. The Berry phase evolves by $2\pi C$ when parallel transported by a reciprocal lattice vector. By contrast, in a RTP the Berry phase evolves by an integer multiple of $2\pi$ when parallel transported for the first half a reciprocal lattice vector, and then unwinds back to zero when parallel transported along the second half. 

\begin{figure}
\includegraphics[width=8cm]{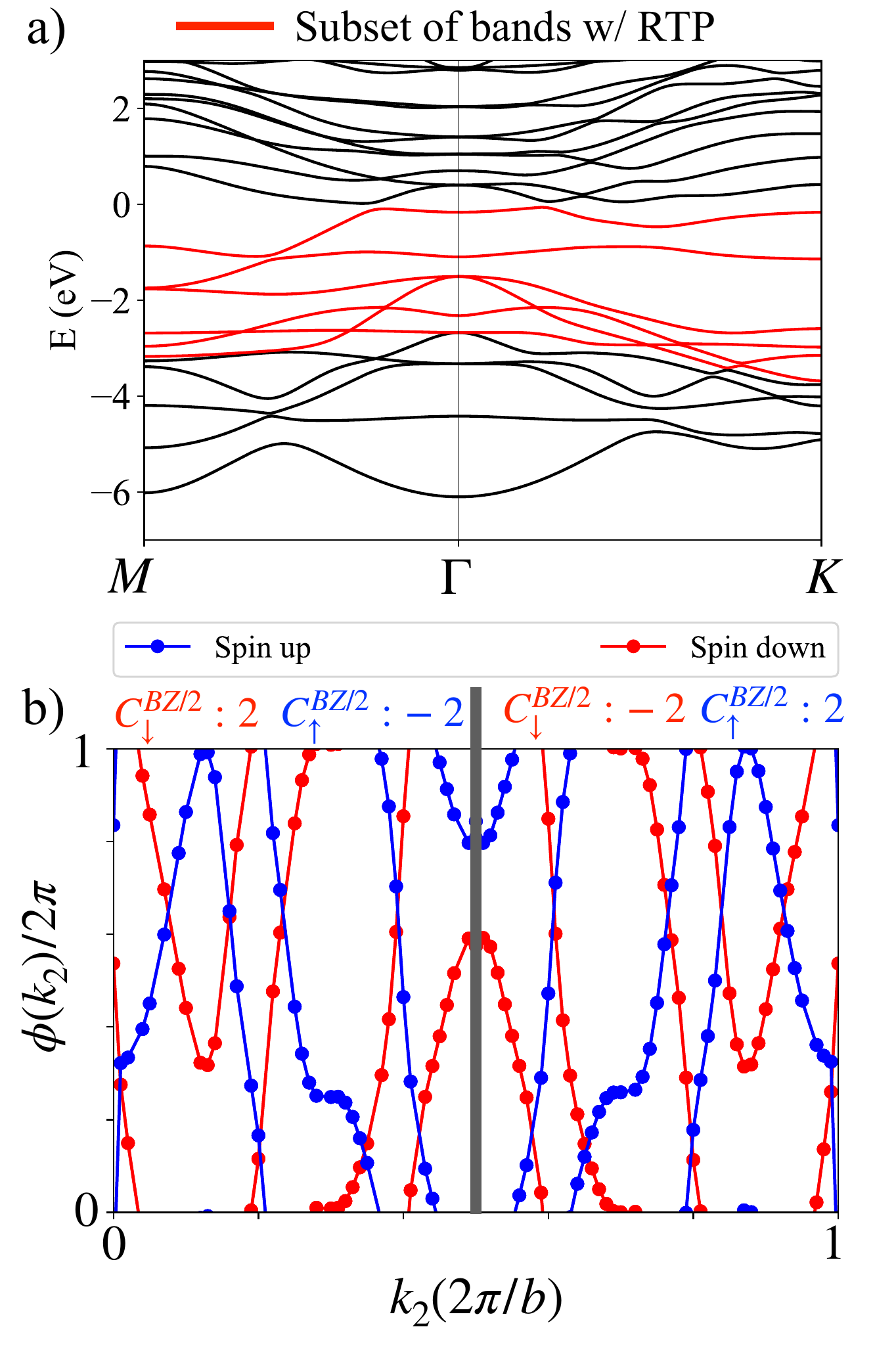}
\caption{(a) Band structure of Ti$_{4}$C$_{3}$ omitting spin-orbit coupling. (b) Results of Berry phase following eq. \eqref{eq:Berry}, considering red bands in (a) as occupied. By decoupling the lower lying trivial bands a quantized RTP appears with a doubled invariant in both the spin-up and spin-down sectors. }
\label{fig:fig5}
\end{figure}

This invariant is fragile to the addition of trivial bands making identification of an exactly quantized RTP exceedingly difficult in material candidates. In three-dimensions it has been used as a means by which to determine the bulk invariant of Hopf insulators and dipolar semimetals\cite{PhysRevLett.101.186805,PhysRevLett.126.216404,PhysRevB.103.045107,PhysRevB.106.075124,PhysRevB.107.115159,PhysRevB.109.L081101,PhysRevB.110.L121122,wang2023fundamentals}. Evidence of an RTP which has been perturbed by realistic considerations is visible in Fig. \eqref{fig:fig2}(a). We note that the region shaded in green corresponds precisely to an RTP. To gain further insight into whether Ti$_{4}$C$_{3}$ supports a quantized RTP which has been trivialized by the coupling of trivial bands, we remove deeper lying occupied bands from the sum in eq. \eqref{eq:Berry} and recompute the Berry flux. Considering only the six occupied bands nearest to the Fermi energy, we find the result shown in Fig. \eqref{fig:fig5}. These results reveal an RTP as the Brillouin zone can be split in half with each half supporting Chern number of equal magnitude and opposite sign quantized to a magnitude of $|\mathcal{C}_{BZ/2}|=|2\pm0.002|$. Fascinatingly, the RTP supports a doubled invariant upon decoupling the lower-lying bands. The observation of an RTP and confirmation of non-trivial topology which has been vanquished by coupling of trivial bands, is also evident in the edge states. In Fig.  \eqref{fig:fig2}(b)-(d) we note the presence of mid-gap states for both the spin up and spin-down sectors. The observed spectra once again points to the existence of non-trivial band topology beyond the paradigm of the 10-fold way\cite{ryu2010topological,RevModPhys.88.035005}.
\section{Discussion}
In this work we have explored a novel two-dimensional antiferromagnetic compound, Ti$_{4}$C$_{3}$, falling within the class of MXenes. This class of materials has proven to be particularly fruitful for both theorists and experimentalists alike due to the presence of both linear-transport due to non-trivial bulk topology\cite{PhysRevB.92.075436,si2016large}, as well as non-linear optical response\cite{gao2020ultrafast,jiang2018broadband}. The antiferromagnetic ordering in Ti$_{4}$C$_{3}$ opens further avenues to explore non-linear response due to the role of magnetic ordering in breaking of centrosymmetry. Similar physics has been identified in the context of MnBi$_{2}$Te$_{4}$ to great success\cite{wang2023quantum}. Our results thus promote further investigation of the shift current in materials for which the magnetic ordering breaks inversion. 
\par 
Finally, the predicted shift current response is large both in the infrared and visible regimes, motivating future work towards experimental verification and potential utilization. The evidence of fragile topology is an intriguing aspect of this material, and motivates further study of the bulk three-dimensional system and the wider class of AFM MXenes.  
\acknowledgments{} 
 Work at NORDITA is supported by NordForsk. The computations were enabled by resources provided by the National Academic Infrastructure for Supercomputing in Sweden (NAISS), partially funded by the Swedish Research Council through grant agreement no. 2022-06725 and 2023-03894. We also acknowledges financial support from the Knut och Alice Wallenberg Foundation, Sweden, and Kempe-Stiftelserna, Sweden.
\bibliography{references}
\end{document}